\documentclass[12pt]{article} 
\usepackage{natbib}
\usepackage[pdftex]{graphicx}
\usepackage{amsmath}
\usepackage{amsthm}
\usepackage{color}
\usepackage{enumerate}
\usepackage{subfigure}
\usepackage[english]{babel}
\usepackage[utf8]{inputenc} 
\usepackage{booktabs} 
\usepackage{array} 
\usepackage{paralist} 
\usepackage{verbatim} 
\usepackage{subfig} 

\usepackage{etoolbox}

\BeforeBeginEnvironment{figure}{\vskip-1.5ex}
\AfterEndEnvironment{figure}{\vskip-2ex}
\newcommand{\blind}{1}
\newcommand{\bd}{ \boldsymbol}

\theoremstyle{definition}

\usepackage{setspace}

\begin{document}

\def\spacingset#1{\renewcommand{\baselinestretch}%
{#1}\small\normalsize} \spacingset{1.5}


\if1\blind
{
  \title{\bf Achieving Parsimony in Bayesian VARs with the Horseshoe Prior}
  \author{Lendie Follett\thanks{
  The authors thank Cooperative Agreement No. 68-3A75-4-122 between the
USDA Natural Resources Conservation Service and the Center for Survey Statistics
and Methodology at Iowa State University.}\hspace{.2cm}\\
    Department of Data Analytics, Drake University\\
    and \\
    Cindy Yu\\
    Department of Statistics, Iowa State University}
  \maketitle
} \fi

\if0\blind
{
  \bigskip
  \bigskip
  \bigskip
  \begin{center}
    {\LARGE\bf Achieving Parsimony in Bayesian VARs with the Horseshoe Prior}
\end{center}
  \medskip
} \fi

\bigskip
\begin{abstract}
In the context of a vector autoregression (VAR) model, or any multivariate regression model, the number of relevant predictors may be small relative to the information set available from which to build a prediction equation.
It is well known that forecasts based off of (un-penalized) least squares estimates can overfit the data and lead to poor predictions. Since the Minnesota prior was proposed (\cite{doan1984forecasting}), there have been many methods developed aiming at improving prediction performance. In this paper we propose the horseshoe prior (\cite{carvalho2010horseshoe}, \cite{carvalho2009handling}) in the context of a Bayesian VAR. The horseshoe prior is a unique shrinkage prior scheme in that it shrinks irrelevant signals rigorously to 0 while allowing large signals to remain large and practically unshrunk. In an empirical study, we show that the horseshoe prior competes favorably with shrinkage schemes commonly used in Bayesian VAR models as well as with a prior that imposes true sparsity in the coefficient vector. Additionally, we propose the use of particle Gibbs with backwards simulation (Lindsten et al. (2012), Andrieu et al. (2010)) for the estimation of the time-varying volatility parameters. We provide a detailed description of all MCMC methods used in the supplementary material that is available online.
\end{abstract}
\noindent%
\vfill

\newpage
\spacingset{2} 
\section{INTRODUCTION}
\label{sec:intro}

In many macroeconomic forecasting contexts, the number of relevant predictors may be small relative to the information set available from which to build a prediction equation. While we expect to improve prediction accuracy as we grow our information set, it may be necessary to perform some kind of dimension reduction for this to be true (e.g, \cite{bai:2002}, \cite{giannone:2008}, \cite{doz:2011}, \cite{stock:1998}, \cite{Canova:2004}, \cite{bernanke:2004}). Vector autoregression (VAR) models are well liked for their flexibility and rich parameterization. However, it is well known that ordinary least squares (OLS) solutions tend to overfit the data and produce unreliable forecasts. Rearchers have responded to this drawback by proposing many methods aiming at building more parsimonious models and/or intentionally biasing the coefficient estimates towards 0. Shrinkage and sparsity priors are meant to pull the data model towards a simpler null model. In a Bayesian context, shrinkage priors can do this by penalizing the length of the parameter vector while sparsity priors can be used to encourage some elements of the parameter vector to be set exactly to 0. 

Shrinkage priors for use in VAR models have been widely researched since the introduction of the Minnesota prior of \cite{doan1984forecasting}. Research has been done on how to modify the Minnesota prior in order to systematically choose the amount of overall shrinkage applied to the coefficients (\cite{banbura:2010}). This idea is that the amount of shrinkage should increase as the number of predictors increases. This is motivated by the theory introduced in \cite{demol:2008} which proved that, if shrinkage is applied appropriately, predictors resulting from Bayesian regression are consistent. Additionally, \cite{giannone2010prior} introduced a fully-Bayesian methodology for choosing the amount of shrinkage. This idea differs from other shrinkage ideas because, rather than setting them to a fixed value, the shrinkage parameters are treated as unknown and assigned probability distributions. The data then is able to help dictate the appropriate amount of shrinkage. 

Considering a potentially large information set, we know it is unlikely that each series is useful for prediction. However, the difficulty is that we do not know which combination of series is `best' and it is computationally impractical to search through all possible models. \cite{korobilis2013var} develops a Bayesian methodology to perform automatic variable selection using discrete mixture priors. Discrete mixture priors handle sparse situations by imposing positive prior probability at a value of 0 along with a continuous alternative distribution. Via Markov chain Monte Carlo (MCMC), this method provides both parameter estimates as well as the probability of inclusion as it performs what \cite{korobilis2013var} calls `stochastic search' over likely models. This class of models is flexible in prior specification, results in quantities that have a `nice' posterior interpretation relative to shrinkage priors, and seems to perform well in terms of prediction relative to traditional shrinkage priors. The limitation of the discrete mixture model is the computational expense it requires. Though the MCMC only requires a few additional steps relative to the shrinkage priors, these steps can be quite time consuming. 

In \cite{carvalho2009handling}, the horseshoe prior is introduced as a way to handle sparsity much like discrete mixtures, but using computational methods that are more similar to the ones used for shrinkage priors. The horseshoe has the benefit of informative posterior quantities that is not shared by other traditional shrinkage priors. Because of the updating of binary variables that is required for the discrete mixture prior, the MCMC time can be prohibitive for even moderately sized models. With the horseshoe prior, we can obtain posterior quantities that help us understand the strength of interrelationships at no additional computational burden than what is required from the traditional shrinkage priors.

To the best of the authors' knowledge, since the horseshoe prior was proposed, it has never been used in a Bayesian VAR. This paper is one of the first to implement the horseshoe prior in an empirical study to compare the predictive performance against other methods. In this paper, we describe the horseshoe prior, offer our own simulation study, and discuss the relevance of the findings in the context of Bayesian VAR. We conclude by comparing the predictive performance of the discrete mixture prior and the horseshoe prior in a medium sized Bayesian VAR in an empirical study. The focus of this paper is to examine the horseshoe prior in the context of a Bayesian VAR model. However, we also emphasize the importance of modeling stochastic volatility. We describe a random walk process on the log time-varying standard deviations paired with a novel prior scheme for the static correlation matrix. We suggest the use of the flexible and efficient particle Gibbs with backwards simulation as an overall algorithm to sample from all the unknowns involved in a Bayesian VAR. 

We illustrate the in-sample behavior of the horseshoe prior and the discrete mixture prior in simulation studies. First we examine the case where the data is generated under a VAR model where there is true sparsity in the coefficient vector. We show that both prior schemes are able to identify the irrelevant signals. Then we also show that both prior schems are able to identify important signals and allow them to remain large. In each simulation study, we show that a particular posterior summary of the horseshoe prior tends to move in sync with a common posterior summary corresponding to the discrete mixture prior. 

In the empirical study, we model eight macroeconomic and financial series in a Bayesian VAR under various traditional shrinkage schemes, the horseshoe prior, and the discrete mixture prior. We find that the horseshoe prior competes favorably with other shrinkage schemes as well as with the discrete mixture prior. We suggest a graphical presentation of posterior quantities that informs us about the strengths of the interrelationships between the eight series and their lags. 

This paper is organized as follows. Section 2 describes the horseshoe prior of \cite{carvalho2009handling} and compares the shrinkage behavior of the horseshoe prior to that of other traditional shrinkage priors. This section will illustrate the relevance of the horseshoe prior in large regression models, such as large VARs. Section 3 proposes a data model in the form of a VAR with stochastic volatility. We first suggest a formulation of the horseshoe prior in the context of a VAR model. We also describe a prior formulation for a discrete mixture prior and some traditional shrinkage priors in the context of the same VAR model. We then describe a novel approach to the estimation of stochastic volatility and static correlation matrix. We suggest a particle MCMC algorithm for estimation of all unknowns involved in the model and describe this procedure in the appendix. Section 4 details a simulation study that compares shrinkage and elimination patterns of the horseshoe prior and the discrete mixture prior. Finally, Section 5 concludes with an empirical study involving real macroeconomic and financial data where we test the horseshoe prior, the discrete mixture prior, and other traditional shrinkage priors. 

\section{HORSESHOE PRIOR FOR A SPARSE SOLUTION}

Consider the simple means model described in \cite{carvalho2009handling} where $\bd{y} \mid \bd{\mu}, \sigma^2 \sim N(\bd{\mu}, \sigma^2 \bd{I})$. Here, $\bd{y}$ is a $p$-dimensional vector of observed values and $\bd{\mu} = (\mu_1, \hdots, \mu_p)^T$ is a vector of means but we can imagine a situation where the elements of $\bd{\mu}$ are slopes or effect sizes in a regression model. Let us imagine that we have some reason to believe that $\bd{\mu}$ is sparse; that some elements may equal to 0. We can assign a horseshoe prior to $\mu_i$ for $i = 1, \hdots, p$ by letting $\mu_i \mid \lambda_i, \tau  \overset{ind}{\sim} N(0, \lambda_i^2 \tau^2)$ and $\lambda_i  \overset{iid}{\sim} \text{Half Cauchy}(0,1)$.

\begin{figure}[!htb]
\begin{center}
\includegraphics[width=.7\linewidth]{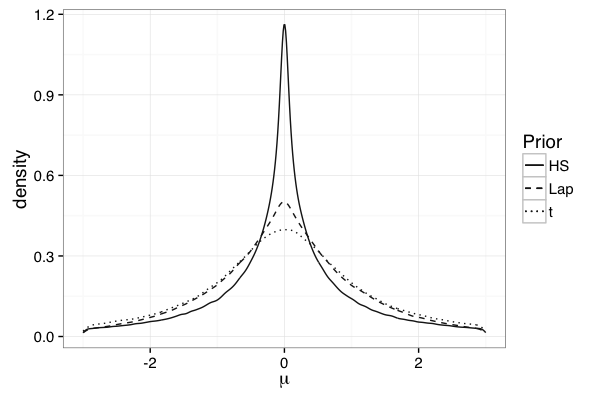}
\caption{\label{horseshoe_hist} Estimated densities of the implied prior on $\mu_i$ based off of the horseshoe prior (HS), student's $t$ prior (t), and the Laplacian prior (Lap). That is, this is an estimate of $f(\mu_i) = \int f(\mu_i \mid \lambda_i)f(\lambda_i)d \lambda_i$. Without loss of generality, we let $\tau$=1.}
\end{center}
\end{figure}

$\tau$ is referred to as the global shrinkage parameter, while $\{\lambda_i: i = 1, \hdots, p\}$ are referred to as the local shrinkage parameters. Here it is not so important what prior is given to $\tau$ - this could also be independently assigned a half-Cauchy distribution. The important part is that $\lambda_i$ has a half-Cauchy distribution. The relatively high density at 0 along with thick tails is what makes the half-Cauchy a special distribution for a shrinkage parameter. The implied prior on $\mu_i$ after integrating out the local shrinkage parameter, $\lambda_i$, is a distribution with relatively fat tails and an infinitely tall spike at 0. Thus, the fat tails will allow $\mu_i$ to become large in the posterior if it needs to, while still rigorously shrinking parameters with small signals. Figure \ref{horseshoe_hist} shows a density of samples taken from this distribution, as well as from the more common t and Laplacian distributions. 

\begin{figure}[!htb]
\begin{center}
\includegraphics[width=.7\linewidth]{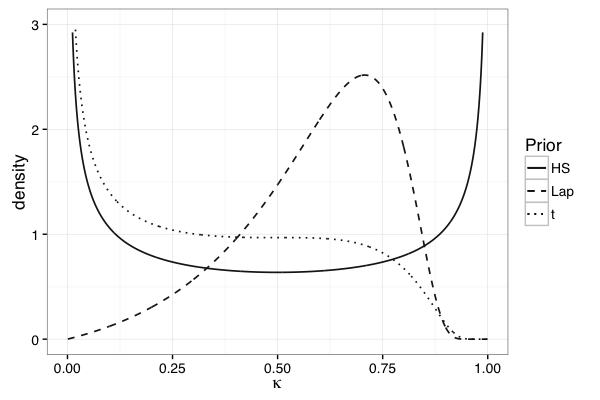}
\caption{\label{kappa_hist} Densities of the implied prior on $\kappa_i$ based on the horseshoe prior, the $t$ prior, and the Laplacian prior.}
\end{center}
\end{figure}

\cite{carvalho2009handling} offer a nice visual of this effect in terms of the posterior weights given to the data and prior mean, 0. For fixed $\tau = \sigma^2 = 1$, they show that $E(\mu_i \mid \bd{y}, \lambda_i^2) = (1-\kappa_i)  y_i + \kappa_i  0 = (1-\kappa_i)  y_i$ where $\kappa_i = 1/(1 + \lambda_i^2)$. The prior on $\kappa_i$, implied by the prior on $\lambda_i$, ends up being a Beta(0.5, 0.5), which is unbounded at 0 and 1 with small mass placed in between (a horseshoe shape). Being unbounded at 0 allows it to let effects grow large, while being unbounded at 1 allows it to shrink effects until they are virtually removed from the prediction equation. Figure \ref{kappa_hist} shows the implied prior on $\kappa_i$ for the horseshoe prior, the student's t prior, and the Laplacian prior. Comparing the horseshoe prior to other popular shrinkage priors such as the ones described in \cite{korobilis2013hierarchical}, we see that for $\kappa_i$ similarly defined, the other priors lead to limited mass at either 0 or 1 or both. The prior on $\kappa_i$ affects the kind of shape that the posterior distribution of $\kappa_i$ is able to take. Limited mass on the extremes of the shrinkage profile limits those priors' abilities to shrink towards 0 and/or to allow the effect to become large. These priors find a shrinkage-compromise between all parameters. As the authors state, this ends up with large signals being over-shrunk and small signals being under-shrunk. In the empirical study, we will show that the distincitve shape of the shrinkage profile corresponding to the horseshoe prior will continue on into the posterior.

The distinction between local and global shrinkage parameters used in the horseshoe prior can be compared with what is done in discrete mixture models. First let $\mu_i = \delta_i \beta_i$ for $i = 1, \hdots, p$.  We can assign a discrete mixture prior for $\mu_i$ by letting $\delta_i \mid \pi  \sim \text{Bernoulli}(1-\pi)$ and $\beta_i \sim f(\beta_i)$ for $i = 1, \hdots, p$.

We write the means as the product of a binary variable, $\delta_i$, and a continuously distributed alternative, $\beta_i$. Here, $\pi$ is the prior probability of the $\{\mu_i: i = 1, \hdots, p\}$ parameters being equal to 0. $\pi$ can be assigned a prior distribution. So, the overall sparsity level for the discrete mixture prior is controlled by $\pi$ where in the horseshoe prior as described above, it is controlled by $\tau$. The discrete mixture prior can let large signals stay large by carefully choosing the form of $f(\beta_i)$. So taking $f$ to be a $t$ distribution, possibly with 1 degree of freedom, which results in a Cauchy prior, may be an effective way to allow large effect sizes to remain large and unshrunk. A discrete mixture prior has the capability of setting coefficients corresponding to irrelevant series exactly to 0, effectively removing them entirely from the predictive equation. In this way, MCMC methods with this prior formulation will result in a search over reasonably likely models.  

Both the discrete mixture prior and the horseshoe prior are defined as hierarchical models so that the optimal degree of sparseness is determined, in part, by the data itself. This way is appealing, as the solution can be thought of as objective and data-driven. 

\section{VAR MODEL WITH SHRINKAGE, SPARSITY, AND STOCHASTIC VOLATILITY}
Let $y_{it}$ represent the $i^{th}$ ($i=1,\hdots, n$) response of interest at time $t$ ($t=1, \hdots, T$). Assuming a maximum lag of $P$, let the data model be written as:
\begin{equation}
\begin{pmatrix} y_{1t} \\ \vdots \\ y_{nt}\end{pmatrix}=\begin{pmatrix} \sum_{j = 1}^n \sum_{k = 1}^P \theta_{1jk} y_{j, t-k}  \\ \vdots  \\  \sum_{j = 1}^n \sum_{k= 1}^P \theta_{njk} y_{j, t-k}  \end{pmatrix} + \begin{pmatrix}\epsilon_{1t} \\ \vdots \\ \epsilon_{nt} \end{pmatrix}. \label{eq:datamodel}
\end{equation}

In the above formulation, $\theta_{ijk}$ represents the effect of the $k^{th}$ lag of the $j^{th}$ series on series $i$. For each of the models we describe below, we expect the presence of stochastic volatility and model it. That is, $\bd{\epsilon}_t \overset{ind}{\sim} N(\bd{0}, \bd{\Omega}_t),$ where $\bd{\epsilon}_t = (\epsilon_{1t}, \hdots, \epsilon_{nt})^T$. In Section 3.1, we describe several shrinkage and sparsity prior formulations in the context of VARs. The focus of this paper is to compare the behavior and predictive performance of the horseshoe prior and discrete mixture prior but we include a few traditional shrinkage priors for further comparison. Section 3.2 describes a strategy for modeling a time varying covariance matrix, $\bd{\Omega}_t$.

\subsection{Shrinkage and Sparsity Priors}

\subsubsection*{Horseshoe prior}

Here we propose a hierarchical model that appllies the horseshoe prior in the context of a Bayesian VAR. We use shrinkage and sparsity to avoid overfitting which leads to poor predictions for the $n$ equations being considered. We want to control, on an equation by equation basis, the amount of global shrinkage. Therefore, our ``global" shrinkage parameters will actually be equation specific, resulting in $n$ global shrinkage parameters. The local shrinkage parameters are defined exactly as in \cite{carvalho2009handling}. We assume the following distributions for $i, j = 1, \hdots, n$ and $k = 1, \hdots, p$: $\theta_{ijk} \mid \lambda_{ijk}, \tau_i  \overset{ind}{\sim} N(0, \lambda_{ijk}^2 \tau_i^2)$, $\lambda_{ijk} \overset{iid}{\sim} \text{Half Cauchy}(0,1)$, and $\tau_i \overset{iid}{\sim} \text{Half Cauchy}(0,1)$.

Therefore, the random variable $\theta_{ijk} \mid \tau_i$ has a prior distribution with an infinitely tall spike at 0 as well as sufficiently thick tails. Define $\kappa_{ijk}^{HS}$ as $\kappa_{ijk}^{HS} = 1/(1 + \lambda^2_{ijk})$. Recall that the prior on $\kappa_{ijk}^{HS}$ looks like the density in Figure \ref{kappa_hist}. In tables and figures, the abbreviation ``HS" will denote this prior specification. There are other ways in which the horseshoe prior could be applied in the context of a Bayesian VAR. In particular, one can specify his or her beliefs about the overall sparsity level by a different specification of the global shrinkage parameter(s). One could even explore a specification that has a Minnesota prior flavor. We do not explore these alternative formulations in this paper. 

\subsubsection*{Discrete mixture prior}
In the same way that we allow for equation specific global shrinkage parameters when using the horseshoe prior above, we will also allow the probability of a coefficient being equal to 0  when using the discrete mixture prior to be equation specific. Assigning a discrete mixture prior in this context involves specifying a mixture of a point mass distribution at 0 along with an absolutely continuous distribution as an alternative to 0. Specification of the continuous distribution will affect both the model fit and performance as well as the role of the point mass indicators, as described in \cite{korobilis2013var}. First, we write $\theta_{ijk} = \delta_{ijk}\beta_{ijk}$ for $i, j = 1, \hdots, n$ and $k = 1, \hdots, p$. Then we assume the following for $i, j = 1, \hdots, n$ and $k = 1, \hdots, p$: , $\delta_{ijk}\mid \pi_i \overset{ind}{\sim} \text{Bernoulli}(1-\pi_i)$, $\pi_i \overset{iid}{\sim} \text{Beta}(a,b)$, and $\beta_{ijk}    \overset{iid}{\sim} N(0, c^2)$.

With this specification, $\pi_i$ represents the probability of a coefficient in equation $i$ being equal to 0. So $\pi_i$ is a measure of the sparseness of the $i^{th}$ equation. We can influence the penalization of dense solutions by setting the prior for $\pi_i$. In an attempt to be less informative, we set $a=b=1$, representing the prior belief that the sparsness probability is uniformly distributed between 0 and 1. We assume independent normal distributions for the prior on $\beta_{ijk}$ where $c^2$ is chosen to be 9, as in \cite{korobilis2013var}. Thus, the $\beta_{ijk}$ parameters are given what is known as as the ridge regression prior. In tables and figures, the abbreviation ``DM" will denote this prior specification.

One option for an alternative specification would be to assign independent $t_v$ distributions to each of the $\beta_{ijk}$ parameters. The motivation behind this might be to allow large signals to remain large as the $t_v$ distribution has relatively fat tails, especially for small $v$. One could set $v=1$ so that the implied prior is actually a Cauchy distribution. It is important to note that, since the the $t$ priors on the coefficients provide relatively high shrinkage around 0, some of the $\delta_{ijk}$ may be partially unidentifiable. That is, small coefficients are already being shrunk towards 0 which results in the likelihood being similar for $\delta_{ijk} = 0$ and $\delta_{ijk}=1$. This does not present a problem for prediction since it is only important that these insignificant coefficients are clustering around 0, but it is important to keep in mind for interpretation purposes. We assume the ridge regression prior to avoid the identifiability issues. 

\cite{korobilis2013var} notes that the discrete mixture prior, as defined above, is too computationally expensive for use in large VAR models. For example, a VAR with hundreds or even dozens of dependent variables may require too much computational time to be of practical use. The Gibbs sampler is used to sample from each conditional posterior. While the discrete mixture prior only requires one additional `chunk' to sample $\{\delta_{ijk}: i,j = 1, \hdots, n, k = 1, \hdots, p\}$, this happens to be a time consuming step. It is for this reason that we feel the horseshoe prior is an important contribution to the Bayesian VAR literature. We can obtain a parsimonious-like effect while adding no more computational burden than that of traditional shrinkage priors.

Consider the VAR model as written in \eqref{eq:datamodel}. We can interpret $\theta_{ijk}$ as the expected change in the conditional expectation of $y_{it}$, $E(y_{it} \mid \bd{\theta})$, for a one unit increase in $y_{j,t-k}$. That is, the interpretation is conditional on fixed values of all other parameters involved in the model. In a model selection context, this clearly presents a problem of how to interpret the posterior point estimates of the coefficients. As the MCMC searches through likely models, coefficients are entering and exiting the data model. A similar phenomenon is happening when we use the horseshoe prior. Thus, we will not report posterior summaries of $\theta_{ijk}$ or $\beta_{ijk}$, but rather the shrinkage and elimination patterns.

\subsubsection*{A selection of traditional shrinkage priors}
The traditional shrinkage priors that we consider in this paper assume the same form where, for $i, j = 1, \hdots, n$ and $k = 1, \hdots, p$, $\theta_{ijk} \mid  \lambda^2_{ijk}  \sim N(0, \lambda_{ijk}^2 )$ and
$\lambda^2_{ijk}  \sim f(\lambda^2_{ijk})$ for some probability density function $f$. In this formulation, $f(\lambda^2_{ijk})$ will control the shrinkage behavior. In general, $f(\lambda^2_{ijk})$ may be conditional on known quantities and/or other unknown parameters to be estimated. We consider four types of distributions for $f()$.

\paragraph{Student's $t$}
We obtain Student's $t$ shrinkage by assuming $\lambda^2_{ijk} \sim \text{Inverse Gamma}(a,b)$.  While setting $a$ and $b$ to small numbers is frequently thought to be noninformative, we set $a=b=1/2$, to obtain a marginal prior on $\theta_{ijk}$ of a $t$-distribution with a location of 0, scale of 1, and 1 degree of freedom. Henceforth, a $t$-distribution with location $a$, scale $b$, and $v$ degrees of freedom will be denoted $t_v(a,b)$. The thick tails of the $t$-distribution allows large signals to get large. However, the $t_v(0,1)$-distribution has more curvature around the origin than a normal and thus shrinks small signals more rigorously. Define $\kappa_{ijk}^{t} = 1/(1 + \lambda^2_{ijk})$. In tables and figures, the abbreviation ``t" will denote this prior specification.


\paragraph{Laplace}
We obtain Laplacian shrinkage by assuming $\lambda^2_{ijk} \sim \text{Exp}(2)$. Define $\kappa_{ijk}^{Lap} = 1/(1 + \lambda^2_{ijk})$. In tables and figures, ``Lap" will denote this prior specification. Posterior means resulting from this formulation agree with the likelihood-maximizing value of Lasso regression \cite{tibshirani1996regression}. 

\paragraph{Ridge} 
Finally, we will test what is probably the most common Bayesian shrinkage method for VARs. This is called the `ridge' prior because the posterior mean of the coefficient corresponds to the solution of ridge regression, \cite{marquardt1975ridge}. This is specified the same way as the continuous part of the discrete mixture model. That is, we set $\lambda_{ijk}^2 = 9$. The predictive performance of the ridge prior with (via our discrete mixture formulation) and without sparsity will be compared in the empirical study.

\subsection{Stochastic Volatility}

Stochastic volatility is generally believed to be present in most financial and macroeconomic data (\cite{clark2014macroeconomic}, \cite{carriero2015common}, \cite{koop2010bayesian}, \cite{karlsson:2012}, among others). Modeling stochastic volatilities will likely result in more realistic prediction intervals as well as result in more efficient estimates of coefficients. \cite{clark2014macroeconomic} consider several different specifications of time-varying volatility and emphasize the importance of considering non-constant variance when using predictive densities. \cite{carriero2015common} propose two stochastic volatility models which involve comovement of the time-changing variances across series.

In the empirical study, each of the prior schemes on the coefficient vector $\bd{\theta}$ will be paired with the same model for stochastic volatility. In the data model of \eqref{eq:datamodel}, the elements of $\bd{\epsilon}_t$ are allowed to be crossectionally correlated but are assumed to be serially independent. The covariance matrix $\bd{\Omega}_t$ is decomposed into a correlation matrix, $\bd{\Psi}$, and time-varying log standard deviations, $\{\omega_{it}: i = 1, \hdots, n \}$. Note, here, that we are assuming the correlation between the errors is constant over time. We can write $\bd{\Omega}_t = diag(e^{\omega_{1t}}, \hdots, e^{\omega_{nt}}) \bd{\Psi}diag(e^{\omega_{1t}}, \hdots, e^{\omega_{nt}})$. The elements of the $n\times n $ matrix $\bd{\Psi}$ are denoted as $\psi_{ii'}$ for $i,i' \in \{1,\hdots, n\}$. That is, $\bd{\epsilon}_t \overset{ind}{\sim} N(\bd{0}, \bd{\Omega}_t)$ where $\bd{\Omega}_{t[i,i']} = exp(2 \omega_{it})$ for $i = i'$ and $\bd{\Omega}_{t[i,i']}=\psi_{ii'}exp(\omega_{it} + \omega_{i't})$ for $i = i$.

For simplicity, we can choose a random walk process for the log time-varying standard deviations by specifying $\omega_{it} = \omega_{i,t-1} + e_{it}$ where  $e_{it} \overset{iid}{\sim}N(0, \tau^2_{\omega i})$. The initial state needs a prior for the Bayesian analysis so we let $\omega_{i 1} \overset{iid}{\sim}N(c,d)$ for $ i=1, \hdots, n.$ The hierarchical model is completed by specifying the following priors: $\bd{\Psi} \sim LKJ(m)$ and 
$\tau_{\omega i} \sim \text{Half Cauchy}(0,1)$ for $i=1, \hdots, n$..

In the above priors, we take $m$, $c$, and $d$ to be known constants. The Lewandowski, Kurowicka and Joe (LKJ) distribution, proposed in \cite{Lewandowski2009}, has become a favorable prior distribution for use in assigning priors to correlation matrices. The LKJ($m$) distribution is a distribution over all possible correlation matrices. If $\bd{\Psi} \sim LKJ(m)$ then $p(\bd{\Psi}) \propto |\bd{\Psi}|^{m - 1}$. These symmetric matrices will have 1 on the diagonal elements and values between -1 and 1 on the off diagonals while maintaining positive-definiteness. The shape parameter $m > 0$ controls the placement of the mass of the distribution. If $m < 1$ then the density of $\bd{\Psi}$ is lowest at the identity matrix. If $m > 1$, then the mode of the distribution is positioned at the identity matrix. We let $m =1$ in our application, which corresponds to a uniform distribution over all positive definite correlation matrices. 

Note that an alternative prior on $\tau^2_{\omega i}$ for $i = 1, \hdots, n$ would be to set $\tau^2_{\omega i} \sim \text{Inverse Gamma}(a,b)$ for  $i = 1, \hdots, n$. The inverse Gamma distribution for variance components is computationally convenient as it lends a closed form for the conditional posterior of $\tau^2_{\omega i}$. An inverse Wishart prior on the covariance matrix would be a more general approach which, while implying inverse Gamma priors on the diagonal elements, would also model nonzero correlations between changes in log volatilities. However, depending on the choice of $a$ and $b$, the inverse gamma distribution will systematically force truly small variances to be over-estimated. The half-Cauchy distribution has positive mass at 0 and does not restrict this area. The half-Cauchy distribution has come to play an important role in hierarchical models (e.g., \cite{gelman2006}). In simulation studies we have observed that, in scenarios where the true variance changes very little from time to time (e.g., constant variance), the inverse Gamma will model fluctuations that are not there while the half-Cauchy results in posterior estimates that are smoother. 

\cite{clark2014macroeconomic} and \cite{carriero2015common} both employ a stochastic volatility estimation method proposed by \cite{kim1998stochastic} which is an accept/reject algorithm. Also in \cite{kim1998stochastic}, an offset mixture of normals is used in a Gibbs algorithm to sample the log volatilities; this method is described in \cite{karlsson:2012}. The authors use transformations and approximations to turn the inherently non-Gaussian and non-linear problem of sampling the volatility parameters into one that is normal and linear. We propose a more general procedure for sampling from the joint posterior distribution of the volatility states and other unknown parameters. This procedure involves particle Gibbs which uses the original state and observation equations. Particle MCMC of \cite{andrieu2010particle}, and in particular, particle Gibbs, combines MCMC and sequential Monte Carlo (SMC) to sample from a potentially high dimensional posterior distribution. It is well suited for large state space models, such as a stochastic volatility model. The algorithm that we employ, proposed by \cite{lindsten2012use}, is called particle Gibbs with backwards simulation. Here, particle Gibbs is paired with a backwards simulation step to improve mixing. 

	Appendix A in the supplementary material briefly describes the algorithm used to sample to volatility parameters $\{\bd{\omega}_t: t = 1, \hdots, T\}$. We've also performed simulation studies in order to evaluate how well this algorithm is able to model variances that change over time. Since our focus is to compare the horseshoe prior and the discrete mixture prior, we do not report on the results here. Details can be seen in the supplemental file. Appendices B and C detail the MCMC algorithm used to sample from the joint posterior distribution based on the horseshoe and discrete mixture priors, respectively.

\section{SIMULATION STUDIES}

In this section we offer three simulation studies meant to illustrate and compare the in-sample behaviors of the discrete mixture prior and the horseshoe prior. Specifically, we focus on how well the models are able to capture true non-zero signals as well as how well they shrink irrelevant coefficients to 0.

In each simulation study below, we generate data according to \eqref{eq:datamodel} where $T=200$, $n=8$, and $P=4$. In total, the dimension of $\bd{\theta}$ is 256. For simplicity, we set $\bd{\Omega}_t = \bd{\Omega} = 0.1\bd{I}_{8\times 8}$ in order to focus on the comparison of posterior coefficient exclusion patterns. For each study, and for both the horseshoe prior and the discrete mixture prior, we run a particle Gibbs with backwards simulation algorithm for 15,000 iterations, discarding the first 5,000, and using the last 10,000 iterations for posterior summaries.  

We focus on the shrinkage profiles in terms of the estimated posterior density of $\kappa_{ijk}^{HS}$ for the horseshoe prior and $\hat{E}(\delta_{ijk} \mid \bd{Y}^T)$, the posterior mean of $\delta_{ijk}$, for the discrete mixture prior. For the horseshoe, we report the maximum a posteriori (MAP) estimate of $\kappa_{ijk}^{HS}$ to give an idea of where most of the posterior mass is located. For the discrete mixture prior, $\hat{E}(\delta_{ijk} \mid \bd{Y}^T)$ can be interpreted as the posterior probability that $\theta_{ijk}$ is in the best model for the $i^{th}$ equation. 

\paragraph{Simulation 1} The purpose of the first simulation is to show the behavior of the discrete mixture prior and horseshoe prior in the context of a VAR where the true coefficient vector contains a given degree of sparsity. The following process is used to simulate the data. For each $i=1, \hdots 8$, $j = 1, \hdots, 8$, and $k = 1, \hdots, 4$, set $
 \theta_{ijk} =\left\{ 
  \begin{array}{l}
0 \text{ with probability 0.5};\\
\text{draw from }N(0,.3^2) \text{ with probability 0.5}. \\
  \end{array} \right.
$

 So, approximately half of the elements of $\bd{\theta}$ are equal to 0 with the rest of the signals being modest in size and independently and identically distributed. In the next two simulation studies, we examine the case where there is no true sparsity in the coefficient vector.

\paragraph{Simulation 2} The vector of coefficients, $\bd{\theta}$, in this sumulation contains all modestly sized signals. There is no true sparsity. The following process is used to generate the data for the second simulation. For each $i=1, \hdots 8$, $j = 1, \hdots, 8$, and $k = 1, \hdots, 4$, simulate $\theta_{ijk} \overset{iid}{\sim} N(0,.15^2).$

\paragraph{Simulation 3} The $\bd{\theta}$ vector in this simulation contains a group of very strong signals mixed with a group of more modestly sized signals. There is no true sparsity. For each $i=1, \hdots 8$, $j = 1, \hdots, 8$, and $k = 1, \hdots, 4$, simulate $
 \theta_{ijk} \text{ from }\left\{ 
  \begin{array}{l}
 N(0,.15^2) \text{ with probability 0.8};\\
 N(.5,.05^2) \text{ with probability 0.1};\\
 N(-.5,.05^2) \text{ with probability 0.1.}\\
  \end{array} \right.
$



\begin{figure}[h!]
\begin{center}
\includegraphics[width=.5\linewidth]{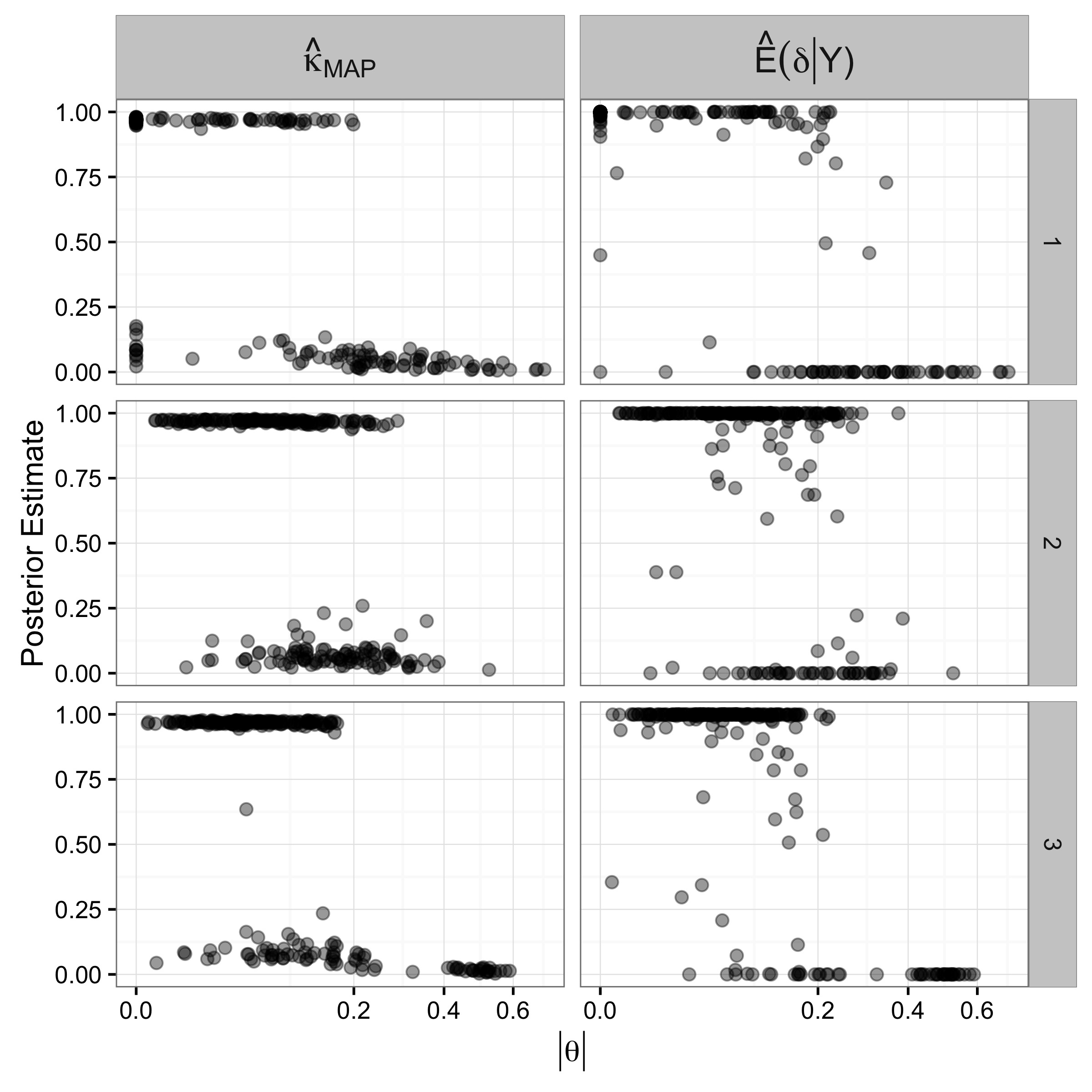}
\caption{\label{sim_shrinkage_summary} Summaries of shrinkage profiles for simulation 1 (top), simulation 2 (center), and simulation 3 (bottom). The $x$-axis represents the absolute value of the true $\theta_{ijk}$ coefficients. The $y$-axis of the left column represents the MAP estimate of $\kappa^{HS}_{ijk}$. The $y$-axis of the right column represents $\hat{E}(\delta_{ijk}\mid \bd{Y}^T)$, the posterior mean of $1-\delta_{ijk}$.}
\end{center}
\end{figure}

Figure \ref{sim_shrinkage_summary} shows the MAP estimate of $\kappa^{HS}_{ijk}$ and the posterior expected value of $1-\delta_{ijk}$ in relation to the absolute value of the true $\bd{\theta}$. Both of these posterior estimates will fall in the range (0,1). Recall that, for $\kappa^{HS}_{ijk}$ close to 0, little shrinkage is being applied while for $\kappa^{HS}_{ijk}$ close to 1, extreme shrinkage is being imposed. In general, for each simulation study, coefficients with truly small values tend to have most $\kappa^{HS}_{ijk}$ posterior mass close to 1. Similarly, those coefficients tend to have a very small posterior mean for $\delta_{ijk}$, meaning a very high posterior probability of exclusion. Visually, we can estimate `cutoffs' for extreme shrinkage or consistent elimination from the prediction equation.

For the first simulation, which involves true sparsity in the coefficient vector, the horseshoe prior typically allows coefficients with absolute values greater than approximately 0.2 to stay large. The discrete mixture prior has high inclusion probabilities for approximately the same range of coefficients. Coefficients with absolute values between 0.1 and 0.2 have similarly varying degrees of shrinkage and elimination for the horseshoe and the discrete mixture prior, respectively. For each of the coefficients with true values exactly equal to zero, the posterior probability of exclusion, $1-\hat{E}(\delta_{ijk}\mid \bd{Y}^T)$, is 1 or close to 1 and the posterior mass of $\kappa_{ijk}^{HS}$ is often close to 1, implying extreme shrinkage. Although it's difficult to see in \ref{sim_shrinkage_summary}, approximately 90\% of coefficients with true values of 0 have a corresponding MAP estimate of $\kappa^{HS}_{ijk}$ that is greater than 0.9. The posterior probability of exclusion is greater than 0.9 for 98\% coefficients in the discrete mixture model. This indicates that both methods effectively remove the irrelevant coefficients when true sparsity exists.

The second simulation study involves data generated from a VAR where all of the coefficients were relatively modest in size, but are not exactly equal to 0. For coefficients with absolute value bigger than 0.2, the discrete mixture results in relatively many consistently eliminated from the model, while the horseshoe seems to allow more coefficients into the predictive equations. Still, we see very similar shrinkage and elimination of coefficients with small true values for both methods.

The third simulation study adds an interesting feature to the data generating process of the second simulation study. Approximately 20\% of the coefficients are far away from the origin relative to the rest of the coefficients. The largest cluster of coefficients are almost always included in the prediction equations for the discrete mixture prior. Similarly, the horseshoe prior places very little shrinkage on these coefficients and allows them to remain large. Relative to the first two simulated data sets, the horseshoe prior sees more MAP estimates in the mid-range between 0 and 1 for modestly sized signals. The discrete mixture prior seems somewhat less likely to eliminate larger signals when large signals are more common. For small true values, both the horseshoe and the discrete mixture tend to eliminate them from the model.

These simulation studies illustrate the shrinkage and sparsity behavior of the horseshoe prior and the discrete mixture prior. In studies 1 and 3 we see that both methods not only are able to recognize and virtually remove irrelevant signals, but that they also are able to detect important signals and leave them unshrunk. In study 2, the horseshoe seems to allow more signals to remain in the predictive equations relative to the discrete mixture prior. Further, we see that the two methods result in posterior summaries that offer similar information concerning the strength of interrelationships. While the resulting predictive equations are similar between the horseshoe prior and the discrete mixture prior, the horseshoe prior results in a MCMC algorithm that has an obvious computational advantage. The time it takes to run the MCMC algorithm based on the horseshoe prior is just a fraction of the time it takes to run the MCMC algorithm based on the discrete mixture. We observe that that horseshoe running time takes, on average, $1/20$ of the discrete mixture running time.

\section{EMPIRICAL STUDY }
The simulation studies illustrate what we may be able to expect in terms of the shrinkage and elimination of coefficients. Now we aim to examine the predictive performance of the six models using real macroeconomic and financial series. That is, we will be examining results from Bayesian VARs with the following priors on the coefficients: horseshoe (HS), discrete mixture (DM), student's t (t), Laplacian (Lap), and ridge. We can compare these models in terms of predictive performance and interpretability.

\subsection{Data}

We have 8 quarterly financial and macroeconomic series from Q1 of 1960 to Q4 of 2010: 1-year treasury constant maturity rate (GS1), real gross domestic product (GDPC96), gross domestic product implicit price deflator (GDPDEF), unemployment rate (UNRATE), total nonfarm payrolls: all employees (PAYEMS), M1 money stock (M1SL), M2 money stock (M2SL), and velocity of M1 money stock (M1V). Series that are collected on a monthly basis were aggregated to be quarterly. We use the same data transformations as described in \cite{korobilis2013hierarchical} to obtain stationarity. Finally, all data series are standardized by their means and variances. All data is obtained from the St. Louis Fed FRED database and downloaded using the R function \texttt{getSymbols} of the package \texttt{quantmod}. We will refer to these series using the FRED series ID. 



The data model is just as in \eqref{eq:datamodel} where we take $P$, the max lag, to be 4. The dimension of $\bd{\theta}$ is 256. This basic form will be paired with each of the previously mentioned shrinkage and sparsity prior schemes for $\bd{\theta}$. Each of these Bayesian models incorporate stochastic volatility as described in Section 3.2. 

\subsection{Posterior summaries}

Particle Gibbs algorithms based on each of the six Bayesian models are run for 30,000 iterations on the a subset of the available data set from 1960:Q1 to 1999:Q1 (T = 150). The first 5,000 iterations are used to tune the tuning parameters for random walk Metropolis steps and discarded as burn-in.
The initial values for $\bd{\theta}$ are set to the OLS solution. The initial values for the volatility parameters, $\{\bd{\omega}_t: t = 1, \hdots, T\}$ are set arbitrarily to $iid$ draws from a N(0,1) distribution. For the discrete mixture prior, we start with $\delta_{ijk} = 1$ for $i,j=1, \hdots, 8$ and $k=1, \hdots, 4$. Other unknowns are set to reasonable constants. 

Given that we are using shrinkage and sparsity as ways to achieve better predictive performance, a meaningful way to compare the behavior of the four models is to compare their posterior ``shrinkage profiles", that is, by looking at the posterior distribution of $\kappa$. While the $\kappa$ parameters cannot directly be interpreted as the posterior weights corresponding to the OLS solution, they still offer an intuitive visual of the way various priors are applying local shrinkage rules. For the shrinkage models, we can compare the posterior distribution of the $\kappa_{ijk}$ parameters as described in the previous section. In this paper, we present the posterior distributions only for the horseshoe prior and focus on comparing this visual to the posterior inclusion probabilities resulting from the discrete mixture prior. We choose to compare posterior summaries of the coefficients involved in the M1SL prediction equation in order to save space.

\begin{center}
\begin{figure}[!htb]
\includegraphics[width=\linewidth]{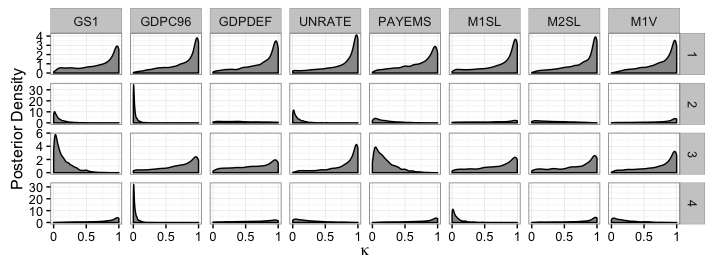}
\caption{\label{shrinkage_profiles_HS} Posterior density of $\kappa^{HS}_{6jk} = 1/(1 + \lambda^2_{ijk})$ using sample from 1960:Q1 to 1999:Q1. That is, these show posterior densities of the shrinkage coefficient involved in the prediction equation for M1SL. Each row shows the densities for the corresponding predictive series ($j$). The columns indicate the different lags ($k$). }
\end{figure}
\end{center}

Recall that the horseshoe prior assumes a prior distribution for $\kappa_{ijk}^{HS}$ that is unbounded at 0 and 1. This leads to distinctive shapes of the posterior densities of $\kappa_{ijk}^{HS}$. Figure \ref{shrinkage_profiles_HS} shows the posterior densities of the $\kappa_{ijk}^{HS}$ parameters involved in the predictive equation for M1SL. Each row of plots corresponds to a lag ($k$) while the columns are predictor-specific ($j$). We note that the scale on the y-axis for the densities varies by each plot. A better representation of the relative importance of a series in an equation would have a constant y-axis scale for each equation. However, to show more detail in the densities, we allow them to vary while being mindful of the differing scales.


A density piled close to 0 indicates an important series while a density piled close to 1 suggests that series is best left out of the prediction equation. Right away, we see none of the first lags seem to be important in predicting M1SL. However, there are many instances where lags 2, 3, or 4 of a series remains unshrunk in the predictive equation. For example, there is evidence that lags 2 and 3 of GS1 and lags 2 and 4 of GDPC96 are included (unshrunk) in the `best' model. Looking at lags of itself, we see that only lag 4 of M1SL is helpful in predicting future values of itself.

For the discrete mixture prior which imposes `true' sparsity, we can first look at the posterior distribution of the $\delta_{ijk}$ parameters. Table \ref{delta_means} shows the posterior means of $\delta_{ijk}$ for the equation corresponding to M1SL. Each column represents  the series corresponding to a coefficient within a predictive equation. Lags of predictive series are shown as rows. The entries of this table are posterior probabilities that the $k^{th}$ lag of the $j^{th}$ predictor is present in the equation predicting M1SL. We gather that some series are always present in a prediction equation while some never are. 

Comparing Table 1 with the posterior shrinkage profiles of the horseshoe prior, Figure \ref{shrinkage_profiles_HS}, we can see that the coefficients that had a high probability of entering the prediction equation for M1SL correspond to a series with a lag of 2,3, or 4. In particular, lags 2 and 3 of GS1 and lags 2 and 4 of GDPC96 are quite prominent in the prediction equation for M1SL. In fact, the posterior inclusion probabilities shown in Table 1 echo what is shown in Figure \ref{shrinkage_profiles_HS} for the posterior shrinkage profiles of the horseshoe prior for almost all series. The exceptions are lag 4 of UNRATE where the horseshoe does not apply shrinkage but the discrete mixture excludes consistently, and lag 2 of M1SL where the horseshoe shrinks the coefficient but the discrete mixture includes consistently.

The posterior distributions of the $\kappa_{ijk}^{HS}$ parameters can be characterized by the tendency pile close to either 0 or 1. For the horseshoe model, this sort of density plot can be very useful to help visualize how local shrinkage rules were applied in the posterior. Using this information, we can infer the importance/strength of interrelationships. In this paper we have focused only on the predictive equation of M1SL. We include a complete graphical representation of the $\kappa_{ijk}^{HS}$, $\delta_{ijk}$, $\kappa_{ijk}^{t}$, and $\kappa_{ijk}^{Lap}$ parameters in the supplementary material.

In conclusion, the shrinkage and sparsity prior formulations offer very much of the same information. Researchers are able to visually summarize important interrelationships using the posterior information from the horseshoe prior. This is done by viewing the distinctive posterior densities of $\kappa_{ijk}^{HS}$. We conclude that the horseshoe and discrete mixture prior seem to result in similar predictive equations. We may expect them to result in similar one-step-ahead predictions. \cite{carvalho2010horseshoe} shows that, in fact, the horseshoe prior and discrete mixture prior do result in similar predictions.

%

\begin{table}[h!]
\caption{\label{delta_means} Posterior means of $ 1-\delta_{ijk}$. This gives the posterior probability of $\beta_{ijk}$ being excluded from the forecast equation using data from 1960:Q1 to 1999:Q1.}
\centering
\begin{tabular}{lrrrrrrrrr}
  \hline
i & k & GS1 & GDPC96 & GDPDEF & UNRATE & PAYEMS & M1SL & M2SL & M1V \\ 
  \hline
  M1SL &    1 & 1.000 & 0.999 & 1.000 & 1.000 & 0.997 & 0.999 & 1.000 & 1.000 \\ 
  M1SL &    2 & 0.000 & 0.000 & 1.000 & 0.000 & 0.000 & 0.000 & 0.000 & 0.999 \\ 
  M1SL &    3 & 0.000 & 0.942 & 0.998 & 1.000 & 0.000 & 0.999 & 0.799 & 0.999 \\ 
  M1SL &    4 & 1.000 & 0.000 & 0.926 & 0.999 & 0.997 & 0.000 & 0.999 & 0.000 \\ 
   \hline
\end{tabular}
\end{table}

\subsection{Predictive performances}
Out-of-sample predictive performance of these six models is assessed based on 50 one-step-ahead predictions. Each of the 50 one-step-ahead predictions are computed using R=1,000 draws from the particle Gibbs algorithm. For example, the first prediction is for 1999:Q2 and is calculated using all data up to 1999:Q1. By adding one time point at a time to the sample, we compute one-step-ahead predictions up until the prediction for 2010:Q4. For each MCMC run, initial values are taken to be posterior means or medians resulting from the last prediction's MCMC. For all six Bayesian models, we use the mean of samples from posterior predictive distribution of $\bd{y}_{t+1} \mid \bd{Y}^t$ as the point forecast for $\bd{y}_{t+1}$. Thus, each model will lend an estimate of $E(\bd{y}_{t+1} \mid \bd{Y}^t)$. For the shrinkage priors, $m \in \{\text{ridge, Lap, t, HS}\}$, the one-step-ahead forecasts are computed as: $\hat{y}_{it \mid t-1, m} = \frac{1}{R}\sum_{r=1}^R (\sum_{j = 1}^n \sum_{k = 1}^P \theta^{(r)}_{ijk} y_{j, t-k})$ where $\theta^{(r)}_{ijk}$ represents the $r^{th}$ MCMC sample of $\theta_{ijk}$.


Similarly, for the discrete mixture prior, the one-step-ahead forecasts are computed as:
$\hat{y}_{it \mid t-1, DM}  = \frac{1}{R}\sum_{r=1}^R (\sum_{j = 1}^n \sum_{k = 1}^P \delta_{ijk}^{(r)}\beta_{ijk}^{(r)}y_{j,t-k}).$
\cite{korobilis2013var} notes that for the discrete mixture prior, this is Bayesian model averaging. 

  We use the root mean squared forecast error (RMSFE) as a measure of prediction accuracy. Let $\hat{y}_{it \mid t-1, m}$ be the posterior predictive mean for the $i^{th}$ series based on model $m$ where $m \in $ $\{$ridge, Lap, t, HS, DM$\}$. Then, $RMSFE(m,i) = \sqrt{ \frac{1}{50}\sum_{t = 151}^{200} (\hat{y}_{it \mid t-1, m} - y_{it} )^2}$.

\begin{table}[h!]
\caption{\label{RMSFE_table} Relative RMSFE for five Bayesian models: ridge, Laplacian, $t$ priors, the horseshoe, and the discrete mixture. Baseline model is a VAR(4) fit with OLS. Smallest RMSFE is bolded for each series (row).}
\centering
\begin{tabular}{rlllll}
  \hline
 & ridge & Lap & t  & HS & DM \\ 
  \hline
GS1          & 0.726 & 0.696 & 0.707   & \textbf{0.594} & 0.633 \\ 
  GDPC96 & 0.947 & 0.899 & 0.900  & 0.873 & \textbf{0.854} \\ 
  GDPDEF & 0.871 & 0.836 & 0.866 & \textbf{0.820} & 0.905 \\ 
  UNRATE & 0.909 & 0.869 & 0.897  & \textbf{0.826} & 0.854 \\ 
  PAYEMS & 0.869 & 0.878 & 0.877 & \textbf{0.854} & 0.861 \\ 
  M1SL      & 1.057 & 1.014 & 1.037 & \textbf{0.884} & 0.932 \\ 
  M2SL     & 1.090 & 1.023 & 1.058 & 0.940 & \textbf{0.930} \\ 
  M1V       & 1.004 & 0.992 & 0.995 & \textbf{0.885} & 0.888 \\ \hline
   average& 0.934 & 0.901 & 0.917 & \textbf{0.834} & 0.857 \\ 
   \hline
\end{tabular}
\end{table}


Table \ref{RMSFE_table} shows the RMSFE reported as ratios relative to a VAR(4) fit with OLS. We choose a lag of 4 for all models because this is common practice for quarterly data. The discrete mixture prior and the horseshoe prior seem to have similar forecast performance. However, based on Table \ref{RMSFE_table}, the horseshoe results in the smallest error measures more often. On average, the horseshoe prior results in the lowest RMSFE. Recall that the continuous part of the discrete mixture prior was specified as the ridge regression prior. By comparing ridge to DM in terms of RMSFE, we see that the ridge prior clearly seems to benefit from the sparsity component. Both Laplacian and ridge regression priors tend to do the worst, but can still out perform OLS. This may be caused, at least in part, by the fact that these Bayesian models incorporate stochastic volatility, which may improve coefficient estimates.


Based on the results from this data set, we can make a recommendation for a `default' prior specification. We consider prediction performance, computational expense, and interpretability. In terms of prediction performance, the horseshoe and discrete mixture priors perform the best in terms of RMSFE, with the horseshoe performing slightly better. The horseshoe prior is relatively easy to implement. Computationally, the discrete mixture prior is disadvantaged relative to the other shrinkage schemes, because of the sampling of $\{\delta_{ijk}: i,j=1,\hdots,n,k=1,\hdots P\}$. However, the discrete mixture prior offers a nice feature of the interpretability of the posterior means of $\delta_{ijk}$, which gives the researcher an idea of importance for each series. We claim that the horseshoe has a similar feature in the posterior densities of $\kappa_{ijk}^{HS}$ which have distinct shapes that tend to move in sync with the posterior means of $\delta_{ijk}$, as illustrated in the simulation study and Section 5.2. The horseshoe prior, via posterior densities of $\kappa_{ijk}^{HS}$, provides a nice way to visualize interdependencies between series in a multivariate regression context. In summary, we have defined three advantage sets; one for prediction performance, one for computational expense, and one for interpretability. The horseshoe prior falls into each of these advantage sets.

\section{CONCLUSIONS}

In this paper we have discussed using a shrinkage prior, the horseshoe prior, as a way to avoid overfitting the data. Although the horseshoe prior is computationally much like a shrinkage prior, the posterior behavior is similar to a true sparsity solution. Allowing for each coefficient to have its own prior standard deviation around 0 which is assigned a half Cauchy distribution results in extreme posterior shrinkage or lack thereof. The resulting predictive equations closely resemble those obtained via Bayesian variable selection using discrete mixture prior. 


\cite{korobilis2013var} showed that, in a forecasting context, sparsity was highly competitive with traditional shrinkage methods. We saw in an empirical study that the predictive performance of the horseshoe prior is very close to that of the discrete mixture priors, and often beats it. Additionally, we want to emphasize that the horseshoe prior is as easy to implement as many traditional shrinkage methods and computationally less expensive than the discrete mixture prior. For these reasons, we feel the horseshoe prior is a valid competitor of traditional shrinkage methods as well as a useful alternative to discrete mixture prior. 

\section{SUPPLEMENTARY MATERIALS}
\begin{description}

\item[Appendix:] PDF file containing derivations of all conditional posteriors used in the particle MCMC algorithm, a detailed description of the conditional auxiliary particle filter, and a simulation study demonstrating the efficiency of the conditional auxiliary particle filter. We also include a full graphical presentation of the posterior shrinkage profiles mentioned in the empirical study as well as the full table of posterior inclusion probabilities.

\item[Rcpp code for Horseshoe particle MCMC:] Rcpp file containing code to perform particle MCMC when coefficients are given the horseshoe prior.

\item[Rcpp code for discrete mixture particle MCMC:] Rcpp file containing code to perform particle MCMC when coefficients are given a discrete mixture prior.


\end{description}

\bibliographystyle{Chicago}

\bibliography{Bibliography-MM-MC}
\end{document}